# Second comment on „Dense and nanometric electronic excitations induced by swift heavy ions in an ionic CaF$_2$ crystal: Evidence for two thresholds of damage creation"

M. Karlušić*

*Ruđer Bošković Institute, Bijenička 54, 10000 Zagreb, Croatia*

Controversy over nature and existence of the velocity effect [G. Szenes, Phys. Rev. B **87**, 056101 (2013)], [M. Toulemonde *et al.*, Phys. Rev. B **87**, 056102 (2013)] reignited after new experimental data on swift heavy ion tracks in CaF$_2$ was reported recently [M. Toulemonde *et al.*, Phys. Rev. B **85**, 054112 (2012)]. Here we show that results of the analysis in ref. [G. Szenes, Phys. Rev. B **87**, 056101 (2013)] should be interpreted differently within analytical thermal spike model. We also propose explanation for apparently missing velocity effect.

Swift heavy ion (SHI) track size depends on its electronic stopping power (S$_e$), as lost kinetic energy of the impinging SHI is deposited in the material along its trajectory. If density of deposited energy is sufficient to induce melting, upon rapid quenching ion track is formed. However, it was observed that ion tracks observed after 10 MeV/u SHI irradiations are typically much smaller than 1 MeV/u SHI tracks, even when electronic stopping power is the same [AM93]. Velocity effect is an important issue at present because two most commonly used thermal spike models that describe formation of SHI tracks provide different interpretation of this effect. Inelastic thermal spike model (ITSM) relates it to different energy deposition profiles due to the velocity of the SHI [AM93], [MT12a], but analytical thermal spike model relates this effect to the contribution from Coulomb explosion at small SHI velocities [GS11]. Resolution of this open question would in turn present major argument for the validity of one model over the other, since models are fundamentally incompatible with each other.

There are several papers related to the SHI tracks in CaF$_2$, so first we give chronological review of experimental and theoretical development. Followed by the TEM observation of cluster ion tracks in CaF$_2$ [JJ98], ATSM analysis concluded that velocity effect is absent, i.e. it was shown that low velocity cluster ions induce tracks that can be described with model parameters for high velocity irradiations [GS00], assuming melting as requirement for ion track formation. In later works, according to TEM observations, it was shown that SHI from high velocity regime induce even smaller tracks [NK05], [SAS07], hence velocity effect was apparently confirmed. Smallness of observed ion tracks was described within ITSM by introducing boiling as requirement for ion track formation, instead of usual melting of the material [MT12a] and identical description was proposed in terms of ATSM in ref. [MK12].

Recently, RBS/c and XRD were used to investigate ion tracks in CaF$_2$ and results were interpreted using ITSM [MT12b]. In that work, RBS/c data was related to the melting of the material since observed ion tracks were much larger than tracks observed by TEM. Similarly, XRD measurements of peak area and peak width were assigned to the boiling and melting, due to good correspondence with the TEM and RBS/c data, respectively. However, as pointed out in Comment [GS13], both RBS/c and XRD linewidth data do not show velocity effect in wide energy range (1 MeV/u – 10 MeV/u). In reply to the Comment [MT13], authors pointed out that velocity effect is clearly visible from TEM measurements, but clear refutation of Comment [GS13] was missing.

---

* marko(dot)karlusic(at)irb(dot)hr

In the following, we confine our analysis to available TEM and RBS/c data. It was proposed in ref. [MK12] that boiling requirement with ATSM parameters $a_0$ = 4.5 nm and g = 0.34 (or g = 0.17) can be used for satisfactorily description of the available TEM data. As shown in Fig.1, agreement between ATSM and experimental data [JJ98], [NK05], [SAS07] is good. We have omitted $Au_4$ data point [JJ98] because of the large nuclear stopping power, and $C_5$ data point [JJ98] because of the very small ion track radii. For 30 MeV $C_{60}$ data point, stopping power is lower then in ref. [JJ98] because it was recalculated using SRIM 2008 [JFZ10].

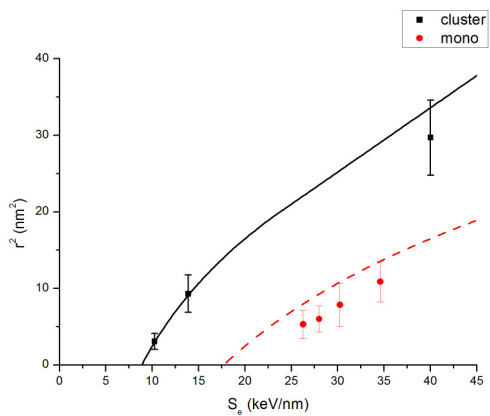

Figure 1. TEM data for ion tracks in $CaF_2$ [JJ98], [NK05], [SAS07] and ATSM prediction with boiling as track formation requirement. Cluster ion track radii can be described with low velocity ATSM parameters ($a_0$ = 4.5 nm, g = 0.34, black full line) and monoatomic ion track radii with high velocity ATSM parameters ($a_0$ = 4.5 nm, g = 0.17, red dashed line).

However, even TEM data show peculiar behaviour related to the velocity effect, as all new SHI track data [NK05], [SAS07] clearly belong to the high velocity regime in Fig. 1. But, according to the ATSM, velocity of 2.4 MeV/u $^{238}$U ions (34.6 keV/nm) should belong to the low velocity regime, and 5.5 MeV/u $^{209}$Bi ions (30.2 keV/nm) should belong to the intermediate velocity regime. Thus, one could speculate that velocity effect is modified, i.e. irradiations using rather slow ions having velocity around 1 MeV/u still belong to the high velocity regime. Such explanation probably could be incorporated within ATSM because shift of the velocity effect (although much smaller) in case of $Y_3Fe_5O_{12}$ and $LiNbO_3$ was seen before, but it could present difficulty for ITSM [GS98]. This in turn could also explain why velocity effect was not observed in RBS/c and XRD data since all data were obtained from irradiations of 1 – 10 MeV/u. On the other hand, cluster ion irradiations with specific ion energy of < 0.2 MeV/u [JJ98] should belong to the low velocity regime as observed with TEM. Therefore, RBS/c analysis of ion tracks formed using ions with specific energy around 0.1 MeV/u would be needed to check this hypothesis.

In Comment, all RBS/c and linewidth XRD data were assigned to the high velocity regime (with melting as track formation requirement), marked with dashed line in Fig. 1 in ref. [GS13], where dotted line represented missing low velocity regime data points. However, more careful quantitative analysis of available experimental data reveals problems with such description. Indeed, one can see that in old ATSM analysis [GS00], cluster ion track TEM data was already assigned to the high velocity regime and these data points are significantly bellow high velocity line from Fig 1 in ref. [GS13]. Actually, linear fit of the RBS/c data [MT12b] from the linear regime [GS98] yields value of parameter g (responsible for description of the velocity effect within ATSM) equal to the value for the low velocity irradiations, i.e. g = 0.34, by imposing melting of the material as ion track forming requirement. Although this value is lower than standard low velocity value of g = 0.4, it is equal in value to our previous analysis of TEM data using boiling as track forming requirement [MK12]. This is important result because absence of the velocity effect in $CaF_2$ proposed in ref. [GS13] is meaningfull only if all the available data can be described using only ATSM high velocity parameter g = 0.17, thus assigning ion track formation mechanism exclusively to the thermal spike [GS11]. Even more surprisingly, RBS/c and TEM data for monoatomic ions, to which melting (RBS/c) and boiling requirements (TEM) have to be applied, can be described only with ATSM parameters for low (RBS/c) and high velocity (TEM) irradiations, although both belong to the same 1 - 10 MeV/u energy range.

To progress further, we re-examined published experimental RBS/c data in [MT12b]. To take into account track overlap, analysis of the RBS/c measurements (i.e. variation of disorder fraction $F_d$ with fluence $\Phi$) was done by fitting data to the Poisson law:

$$F_d = \alpha\left(1 - e^{-A\Phi}\right), \qquad (1)$$

where $\alpha$ represents saturation value and $A$ ion track cross section. This can be linearized for small fluences (the non-overlapping regime):

$$F_d = \alpha A \Phi, \qquad (2)$$

and $\alpha A$ as initial disordering $I_d$ was also reported in [MT12b]. Saturation $\alpha$ indicates that some sort of the $CaF_2$ recovery takes place. Assuming cyllindrical symetry with disordered region located along ion trajectory, from eq. (2) the radius of the ion track (i.e. the disordered region) should be given by:

$$r = \sqrt{\frac{\alpha A}{\pi}}, \qquad (3)$$

while the radius of the excited region after passage of the ion (i.e. region where recovery occurs) is:

$$R = \sqrt{\frac{A}{\pi}}. \qquad (4)$$

Since in refs. [MT12b] and [GS13] only R from eq. (4) was considered, on Fig. 2 we show variation of both $R^2$ and $r^2$ as a function of the electronic stopping power obtained from the RBS/c [MT12b]. Clearly, ion track radii redefined by eq. (3) can be described using melting as track forming requirement and ATSM parameters for the high velocity regime ($a_0$ = 4.5 nm, g = 0.17). This way both RBS/c and TEM monoatomic track data (1 - 10 MeV/u) can be described with high velocity parameters of the ATSM. However, RBS/c data described by eq. (4) should not be dismissed as unphysical, because of the good agreement with XRD linewidth data. Furthermore, both swelling and surface nanostructures due to single SHI impacts show threshold behavior around 5 keV/nm that can be assigned to the melting and low velocity parameters of ATSM [MK12].

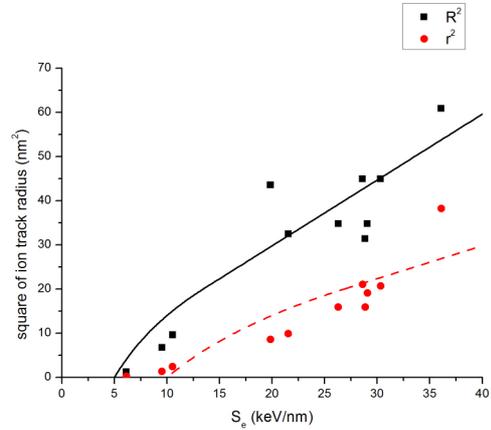

Figure 2. RBS/c data for ion track radii in $CaF_2$ [MT12b], according to definitions given by eqs. (3) and (4). With melting as track forming requirement, shown are predictions of the ATSM using low velocity model parameters ($a_0$ = 4.5 nm, g = 0.34, black full line) and high velocity model parameters ($a_0$ = 4.5 nm, g = 0.17, red dashed line).

In conclusion, we have demonstrated that ion tracks observed with TEM can be described within ATSM using boiling requirement like in the ITSM description. It was also shown that assignment of the RBS/c data to the high velocity regime in ref. [GS13] was incorrect, but reinterpretation of the RBS/c data from [MT12b] made such assignment possible. Explanation for the missing velocity effect was proposed to be related to the shift of the effect to the energies bellow 1 MeV/u.

However, number of important questions remain open:

- why TEM data should be assigned to the boiling of the material?
- why the same RBS/c data (Fig. 2) can be related both to the low and high velocity paramters of the ATSM?
- can ITSM give satisfactory description of the ion track radii according to eq. (3)?
- why are the saturation values different for RBS/c and XRD?

To answer some of those questions, RBS/c analysis of the ion tracks in the energy range around 0.1 MeV/u is needed. Cluster ion irradiations like in ref. [JJ98] or monoatomic ion irradiations (for example 20 MeV iodine) would be most welcome. Re-examining already published data on materials that show similar damage kinetics (RBS/c saturation $\alpha < 1$) like $Al_2O_3$ along the lines described above could also shed some light on this topic.